# Influence of microstructure on the application of Ni-Mn-In Heusler compounds for multicaloric cooling using magnetic field and uniaxial stress

Lukas Pfeuffer [a,*], Adrià Gràcia-Condal [b], Tino Gottschall [c], David Koch[a], Tom Faske[a], Enrico Bruder [a], Jonas Lemke [a], Andreas Taubel [a], Semih Ener [a], Franziska Scheibel [a], Karsten Durst [a], Konstantin P. Skokov [a], Lluís Mañosa [b], Antoni Planes [b] and Oliver Gutfleisch [a]

[a] Institute of Materials Science, Technical University of Darmstadt, 64287 Darmstadt, Germany
[b] Departament de Física de la Matèria Condensada, Facultat de Física, Universitat de Barcelona, 08028 Barcelona, Catalonia, Spain
[c] Dresden High Magnetic Field Laboratory (HLD-EMFL), Helmholtz-Zentrum Dresden-Rossendorf, 01328 Dresden, Germany

**Abstract**

Novel multicaloric cooling utilizing the giant caloric response of Ni-Mn-based metamagnetic shape-memory alloys to different external stimuli such as magnetic field, uniaxial stress and hydrostatic pressure is a promising candidate for energy-efficient and environmentally-friendly refrigeration. However, the role of microstructure when several external fields are applied simultaneously or sequentially has been scarcely discussed in literature. Here, we synthesized ternary Ni-Mn-In alloys by suction casting and arc melting and analyzed the microstructural influence on the response to magnetic fields and uniaxial stress. By combining SEM-EBSD and stress-strain data, a significant effect of texture on the stress-induced martensitic transformation is revealed. It is shown that a <001> texture can strongly reduce the critical transformation stresses. The effect of grain size on the material failure is demonstrated and its influence on the magnetic-field-induced transformation dynamics is investigated. Temperature-stress and temperature-magnetic field phase diagrams are established and single caloric performances are characterized in terms of $\Delta s_T$ and $\Delta T_{ad}$. The cyclic $\Delta T_{ad}$ values are compared to the ones achieved in the multicaloric exploiting-hysteresis cycle. It turns out that a suction-cast microstructure and the combination of both stimuli enables outstanding caloric effects in moderate external fields which can significantly exceed the single caloric performances. In particular for Ni-Mn-In, the maximum cyclic effect in magnetic fields of 1.9 T is increased by more than 200 % to -4.1 K when a moderate sequential stress of 55 MPa is applied. Our results illustrate the crucial role of microstructure for multicaloric cooling using Ni-Mn-based metamagnetic shape-memory alloys.

## 1 Introduction

The continuous growth of global prosperity comes along with a rising demand for energy-efficient and environmentally-friendly refrigeration technologies [1]. It is assumed that the residential energy required for cooling exceeds the energy needed for heating already within the next 50 years [2]. Nowadays, cooling is still based on the vapor-compression technology which utilizes hazardous, explosive or greenhouse gases with up to a thousand times worse global-warming potentials than $CO_2$ [3]. Solid-state caloric cooling is the most promising candidate to address these problems [4,5].

This technology is based on an adiabatic temperature change $\Delta T_{ad}$ or isothermal entropy change $\Delta s_T$ of a ferroic material when subjected to an external stimulus [6]. According to the stimulus used, magnetocaloric (magnetic field), electrocaloric (electric field), barocaloric (hydrostatic pressure), and elastocaloric (uniaxial load) cooling can be distinguished. They all have in common that giant caloric effects with large $\Delta T_{ad}$ and $\Delta s_T$ occur in the vicinity of first-order transformations as a result of the absorbed/released latent heat. However, the inherent hysteresis of first-order transformations limits the materials performance in cyclic operation due to energy losses and/or irreversibilities of the caloric effect [7,8]. As a potential solution, the combination of two external stimuli has been suggested which is referred to as multicaloric effect [9,10].

* Corresponding author:
  E-mail address: lukas.pfeuffer@tu-darmstadt.de (L.Pfeuffer)

The response to multiple stimuli requires a strong coupling of the ferroic order parameters. While a magnetoelectric coupling is rare and often not very pronounced, the opposite is the case for magnetoelastic materials such as La-Fe-Si [11], Fe-Rh [12], Gd-Si-Ge [13] and metamagnetic shape-memory alloys, in particular Ni-Mn-based Heusler compounds [14–16]. Especially the martensitic transformation in the latter exhibits a high sensitivity to both, magnetic and mechanical (uniaxial load and hydrostatic pressure) fields. Thereby, the magnetic field favors the high-temperature, high-magnetization austenite whereas uniaxial load and hydrostatic pressure stabilize the low-temperature, low-magnetization martensite due to its lower crystal symmetry and volume.

Several studies of Ni-Mn-based Heusler compounds show that cyclic caloric effects can be largely tuned by a simultaneous or sequential application of magnetic and mechanical fields [17–19]. In a recent work, we demonstrated that the giant magnetocaloric effect in Ni-Mn-In can be doubled for moderate field changes of 1 T magnetic field when at the same time a stress of 40 MPa is removed [20]. Besides that, a drastically improved cyclic performance was found when a stress is applied during field removal. Similar results were presented for the addition of hydrostatic pressure when the magnetic field is taken off in isostructural Ni-Mn-In-Co [9]. A different strategy is to utilize the thermal hysteresis of first-order magnetoelastic materials rather than to minimize it [21]. In this so-called "exploiting-hysteresis cycle" the material is locked in its ferromagnetic austenite state after magnetic-field application and removal due to the thermal hysteresis. Subsequently, the heat is transferred and the back transformation to the original non-magnetic martensite state is ensured by uniaxial load. In comparison to conventional magnetocaloric cooling, the magnetic field does not have to be applied during the heat transfer in the exploiting-hysteresis concept which allows a significant reduction of expensive permanent magnets as well as the utilization of higher magnetic fields and field-sweep rates.

Although theoretical and experimental studies demonstrated the large potential of multicaloric cooling concepts in Ni-Mn-based Heusler compounds, the influence of microstructure has not been investigated so far. For the single caloric (magneto- and elastocaloric) effects a strong impact of microstructural features such as grain size, grain orientation, phase purity and distribution has been reported [22–25]. However, the use of multiple stimuli requires a tailored microstructure for the response to several external fields and possibly a rethinking of its design strategy.

This work sets out to analyze the influence of microstructure on the response to both, magnetic field and uniaxial load in terms of single caloric effects and multicaloric performance in an exploiting-hysteresis cycle. For that purpose, we selected the Ni-Mn-In Heusler system synthesized by suction casting. While Ni-Mn-In is one of the most promising candidates for multicaloric cooling due to its strong magnetostructural coupling, suction casting provides a facilitated synthesis of common rod- or plate-geometries with good mechanical properties [26–28]. Our results show that the microstructural design has a significant impact on the application of Ni-Mn-based Heusler compounds for multicaloric cooling.

## 2 Experimental details

Ni-Mn-In samples were prepared by manifold arc melting of Ni (99,97 %), Mn (99,99 %), and In (99,99 %). For each sample, a Mn excess of 3 wt% was added to account for evaporation losses during the melting process. The arc-molten ingots were subsequently suction cast into a cylindrical, water-cooled copper mold of 3 mm diameter and 30 mm height, followed by a 24 h annealing at 900 °C under Ar atmosphere and quenching in water.

Temperature-dependent x-ray powder diffraction (XRD) was performed in a purpose-built diffractometer in transmission geometry using Mo K$_\alpha$ radiation, a *MYTHEN R 1K* detector (*Dectris Ltd.*) and a 2θ range from 7 to 57° with a step size of 0.009° [29]. An annealed piece of the arc-molten base ingot was crushed into powder with particle sizes smaller than 40 µm. To release the stresses induced by crushing, the powder was annealed under Ar atmosphere for 7 days at 850 °C and subsequently quenched in water. A *NIST640d* standard reference silicon powder was added to the heat-treated powder to ensure the correction of geometric errors. Afterwards, the mixture was glued on a graphite sheet. The temperature was varied from 310 K to 240 K using a closed-cycle helium cryofurnace (*SHI Cryogenics Group*). XRD data were analyzed by the Rietveld method using *JANA2006* [30] with a superspace approach [31] for modulated martensite structures.

Microstructural characterization by means of electron back-scatter diffraction (EBSD) and chemical analysis via energy dispersive x-ray spectroscopy (EDX) was carried out in a *TESCAN* high-resolution scanning electron microscope. All synthesized samples are in a compositional range of Ni$_{49.8\pm0.1}$Mn$_{35.6\pm0.3}$In$_{14.6\pm0.2}$ and will be denoted in the following as Ni-Mn-In. Temperature-dependent optical microscopy was accomplished at a *Zeiss Axio Imager.D2m* equipped with a LN$_2$ cryostat.

Elastocaloric and mechanical characterization was performed by means of strain-temperature $\varepsilon(T)$ curves under constant load as well as isothermal and quasi-adiabatic stress-strain $\sigma(\varepsilon)$ measurements. Therefore, cylinders of 5 mm height and 3 mm diameter were extracted from the heat-treated rods. The arc-molten samples used for comparison of the stress-induced martensitic transformation behavior and the fracture strength were cut into cuboids with dimensions of 5 mm x 2.5 mm x 2.5 mm. In all elastocaloric and mechanical measurements, the force was applied in compressive mode along the cylinder axis and recorded via a load cell. For the detection of the sample strain, a strain gauge extensometer (*Instron 2620*) was attached to the compression platens [32]. The temperature of the sample was measured by a K-type thermocouple, which was attached to the center of the sample. The $\sigma(\varepsilon)$ measurements were carried out in a universal testing machine *Instron 5967* 30 kN, equipped with a temperature chamber. Thereby, the compressive strength tests and isothermal experiments were performed with a strain rate of 3x10$^{-4}$ s$^{-1}$, while quasi-adiabatic conditions were provided by a strain rate of 3x10$^{-2}$ s$^{-1}$ [33]. The isothermal curves were measured using a discontinuous protocol including heating to the fully austenitic state at 310 K before setting the test temperature. For the determination of $\varepsilon(T)$ under different constant loads, a purpose-built device [34] was used. The compressive load was applied well above the austenite finish temperature $A_f$ to avoid a stress-induced martensitic transformation. Subsequently, the sample was cooled and heated upon the martensitic transition with a rate of 1 Kmin$^{-1}$.

Isofield measurements of magnetization were executed in a *LakeShore* vibrating sample magnetometer (VSM) and a *Quantum Design* physical property measurement system (*PPMS-14 T*) using a heating and cooling rate of 2 Kmin$^{-1}$. For the determination of the isothermal entropy change $\Delta s_T(T,H)$, $M(T)$ curves were recorded in steps of 0.2 T from 0.2 T to 2 T. Additionally, isothermal magnetization measurements up to 10 T were carried out with a sweeping rate of 0.005 Ts$^{-1}$ in temperature steps of 2 K to compute $\Delta s_T(T,H)$ at higher magnetic fields. To erase the materials history after the magnetic field application and removal, a discontinuous measurement protocol was chosen [35]. Thereby, the sample was heated to 15 K above $A_f$ and 50 K below the martensite finish temperature $M_f$ before the measurement temperature was approached. The start and finish temperatures, stresses and magnetic fields of the transformations were evaluated by the double tangent method [36]. The transition entropy change $\Delta s_t$ was determined using a *TA Instruments Q2000* and a *Netzsch DSC 404 F1 Pegasus* differential scanning calorimeter (DSC) with a heating and cooling rate of 5 Kmin$^{-1}$.

Simultaneous adiabatic temperature change $\Delta T_{ad}$ and strain $\varepsilon$ measurements were performed in a purpose-built setup of two nested Halbach magnets [37] and in a solenoid magnet [38] at the Dresden High Magnetic Field Laboratory (HLD). While in the Halbach setup a sinusoidal magnetic field of up to 1.9 T was applied with a highest field-sweep rate of 1 Ts$^{-1}$, magnetic fields of 2, 5, and 10 T were reached in the solenoid with maximum field rates of 318, 866, and 1850 Ts$^{-1}$, respectively. To ensure the comparability of the measurements in both setups, the same specimen including the attached thermocouple and strain gauge was used. For the strain detection, the suction-cast rod was cut along the cylinder axis to obtain a large flat sample surface. Subsequently, one of the half cylinders was separated into two parts parallel to its semicircular plane. The strain measurement was carried out by gluing a linear pattern strain gauge of 0.79 mm gauge length and 1.57 mm grid width to the flat cutting surface of one part of the half cylinder. For the determination of the strain gauge's electrical resistance, a Wheatstone bridge in the Halbach, and a function generator plus a digital lock-in technique in the solenoid setup was used. The gauge direction was perpendicular to the applied magnetic field in both devices. $\Delta T_{ad}$ measurements were performed via a differential T-type thermocouple of 25 μm single wire thickness, which was fixed between the two parts of the half cylinder by a conductive epoxy. In coincidence with the determination of $\Delta s_T(T,H)$, all simultaneous measurements of $\Delta T_{ad}$ were carried out with the abovementioned discontinuous temperature protocol. The temperature of the sample before field application was determined by a Pt100 temperature sensor in both setups. Based on $\Delta T_{ad}$ measurements at the Curie temperature of the austenite $T_C^A$, a starting temperature correction was done for the data from the Halbach setup.

## 3 Results and discussion

### 3.1. Structure analysis

Figure 1 displays the XRD patterns of the annealed Ni-Mn-In powder at 300, 280, and 260 K upon cooling. The diffractograms at high temperature reveal a L2$_1$ structure of the austenite phase which on cooling gradually transforms to low-symmetry 3M modulated monoclinic martensite, being indicated by the changes of the L2$_1$ (111), L2$_1$ (220) and 3M (1210) intensities. The 3M modulation could also be designated as 6M [39]. The corresponding lattice parameters of the full temperature range (310 K – 240 K) are presented in Table 1. Note that the transition width is significantly enlarged in the powder compared to the bulk state of the material (see Figure 4(a)) which is in agreement with the observations for Ni-Mn-In-Co and other first-order magnetocaloric materials such as La(Fe,Si)$_{13}$ and Fe$_2$P-based compounds [40]. Thus in the <40 μm powder, the martensitic transformation from pure austenite at 310 K to an almost entirely martensitic state at 240 K takes place within approximately 70 K, whereas in the bulk sample is reduced to 10 K.

Crystallographically, the conversion from the parent L2$_1$ phase to the 3M modulated monoclinic martensite is described by a linear transformation **F=RU**, that can be separated into a rotation matrix **R** and a transformation stretch matrix **U**. As we are interested in the transformation strains $\varepsilon_{tr}$, solely **U** is considered which has been calculated from the lattice parameters of both phases according to [41,42]. It should be emphasized, that in total 12 transformation stretch matrices exist for the symmetry relation cubic to monoclinic, though their eigenvalues $\lambda_1 \leq \lambda_2 \leq \lambda_3$ coincide. The calculated values for $\lambda_1$, $\lambda_2$ and $\lambda_3$ are given in Table 1 for different temperatures and will be discussed exemplarily for 280 K. ($\lambda_3$-1) refers to the maximum recoverable $\varepsilon_{tr}$ under tension, while (1-$\lambda_1$) characterizes the corresponding value under compression [43]. Hence for our Ni-Mn-In sample, theoretical maximum transformation strains of 5.09 % and 5.79 % are obtained for the application of a tensile and a compressive load at 280 K, respectively. The deviation of the middle eigenvalue $\lambda_2$ from unity $|\lambda_2-1|$ is a measure for the geometrical compatibility of both phases and a key parameter for the width of the thermal hysteresis. *Zhang et al.* [44] demonstrated that the latter significantly increases upon rising values for $|\lambda_2-1|$. In our sample, a small deviation of 0.35 % from $\lambda_2$=1 is observed at 280 K which is outlined by recent reports for Ni-Mn-In of similar transition temperature [42].

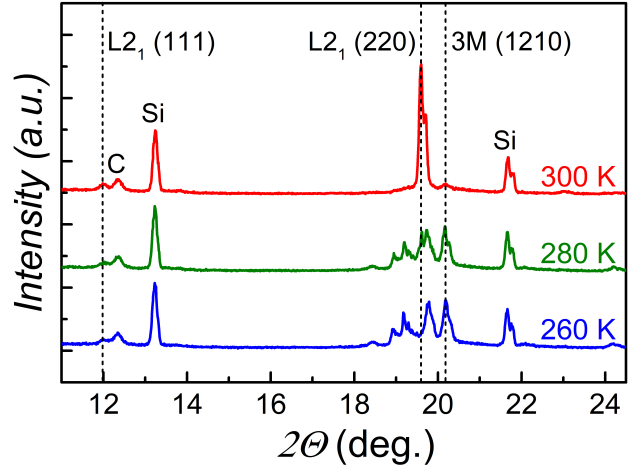

*Figure 1: XRD patterns of the annealed Ni-Mn-In powder at 300, 280, and 260 K recorded upon cooling in zero field. The indexed Bragg peaks indicate representatively the phase transformation from high-temperature $L2_1$ austenite to a low-temperature 3M modulated monoclinic martensite. Si was used as reference material and the C reflection results from the graphite substrate.*

*Table 1: Temperature-dependent evolution of lattice parameters (q represents the modulation vector), eigenvalues ($\lambda_1$, $\lambda_2$, $\lambda_3$), and total volume change $\Delta V/V_0$ of the Ni-Mn-In powder upon cooling from 310 to 240 K. At 310 K, the powder is fully austenitic, while at 240 K it has almost completely transformed to martensite. The XRD patterns and the corresponding lattice parameters at 240 and 310 K were already shown in [45].*

| Temperature | Lattice Parameters | | | | | | $\lambda_1$ | $\lambda_2$ | $\lambda_3$ | $\Delta V/V_0$ |
|---|---|---|---|---|---|---|---|---|---|---|
| [K] | $a_A$ [Å] | $a_M$ [Å] | $b_M$ [Å] | $c_M$ [Å] | $\beta$ [°] | $q$ [Å$^{-1}$] | | | | [%] |
| 310 | 6.0027 | - | - | - | - | - | - | - | - | - |
| 300 | 6.0015 | 4.3879 | 5.6553 | 4.3320 | 92.52 | 0.3489c* | 0.9423 | 1.0037 | 1.0507 | -0.12 |
| 280 | 6.0010 | 4.3881 | 5.6536 | 4.3315 | 92.53 | 0.3505c* | 0.9421 | 1.0035 | 1.0509 | -0.64 |
| 260 | 5.9973 | 4.3920 | 5.6406 | 4.3333 | 92.59 | 0.3516c* | 0.9405 | 1.0042 | 1.0528 | -0.81 |
| 240 | 5.9971 | 4.3947 | 5.6330 | 4.3307 | 92.69 | 0.3535c* | 0.9393 | 1.0032 | 1.0538 | -0.97 |

The transformation stretch matrices can be utilized as well for the calculation of orientation-dependent transformation strains under tension and compression, according to the procedure described in [46]. The computed values for compressive loading along certain crystallographic directions at 280 K are depicted in round brackets in Figures 2(c), 2(f) and 2(i). The temperature of 280 K was chosen as it provides good comparability with experimental conditions and a sufficient martensite phase fraction for a reliable refinement of the lattice constants. In any case, the computed transformation strains exhibit a weak dependence on the temperatures in Table 1, with a maximum deviation of less than 5 % from the values at 280 K. The topmost $\varepsilon_{tr}$ of 5.8 %, corresponding to (1- $\lambda_1$), can be observed when the unit cell of Ni-Mn-In is compressed in [001] direction. For comparison, $\varepsilon_{tr}$ is reduced to 2.6 % when the compressive load is applied along the [101] direction. Therefore, the stress-induced martensitic transformation can be significantly influenced by the texture of the Ni-Mn-In alloy which is discussed in section 3.2. Besides that, the martensitic transformation is accompanied by an overall volume change $\Delta V/V_0$ of -0.97 %. The temperature-dependent changes of the volume are displayed in Table 1.

### 3.2. Microstructural characterization

In section 3.1 the significant influence of the grain orientation/texture on the transformation strain $\varepsilon_{tr}$ was demonstrated. Large recoverable values of $\varepsilon_{tr}$ are desirable for elastocaloric applications and the exploiting-hysteresis concept as the stress $\sigma^{Ms}$ to induce the martensitic transformation can be decreased [47,48]. In addition, microstructural features such as grain size, phase purity and distribution are crucial parameters for the elasto-, magneto- and with this the multicaloric performance of the material [7,49].

In Figure 2 different microstructures of Ni-Mn-In are illustrated by inverse pole figure maps and corresponding texture plots. Note that the textures are discussed qualitatively only due to a limited number of grains. Figure 2(a) depicts the microstructure obtained by suction casting of the material into a cylindrical mold. At the edge of the cylinder a fine-grained chill zone of equiaxed austenite grains is found. Towards the center, opposite to the heat transfer in the cylindrical mold, a radial symmetric growth of large columnar austenite grains with 41 ± 8 µm average diameter and up to 1.1 mm length is observed. The texture plots in Figure 2(b) indicate a <001> solidification texture of the columnar grains with respect to the cylindrical axis. Figures 2(d) and 2(g) illustrate microstructures after arc melting of Ni-Mn-In along the longitudinal plane (A3 ⊥ solidification direction) and transversal plane (A3 ∥ solidification direction) which will be denoted as arc-molten I and arc-molten II in the following. In accordance with suction casting, a <001> solidification texture of the columnar austenite grains is found (see Figures 2(e) and (h)). As a consequence, the texture of the suction-cast and arc-molten I sample show a strong coincidence. However, in contrast to suction casting arc melting results in significantly coarsened grains with an average grain diameter of 675 ± 100 µm and up to 1.5 mm length due to a lower cooling rate during solidification.

In order to evaluate the consequences of the <001> solidification texture on the stress-induced martensitic transformation, the grain orientations in compression direction (CD) have to be considered. This is visualized in the inverse pole figure plots in Figure 2(c), 2(f) and 2(i) for the suction-cast sample, arc-molten I and arc-molten II, respectively. It is apparent that the solidification behavior results in preferential grain orientations in [001] direction in arc-molten II. In arc-molten I and the suction-cast sample mainly grain orientations between [001] and [101] are found, while orientations in [111] direction are not observed. As discussed in section 3.1, the unit cell of the Ni-Mn-In compound exhibits a $\varepsilon_{tr}$ of 5.8 % and 2.6 % when compressed in [001] and [101] direction, respectively. Indeed it was observed by stress-strain $\sigma(\varepsilon)$ measurements that $\varepsilon_{tr}$ in [001] direction reaches 5.7 % for directionally solidified Ni-Mn-In [24] which is in good agreement with the predicted value. Hence, for the microstructures depicted in Figure 2 the highest $\varepsilon_{tr}$ and with this the lowest $\sigma^{Ms}$ is expected for arc-molten II.

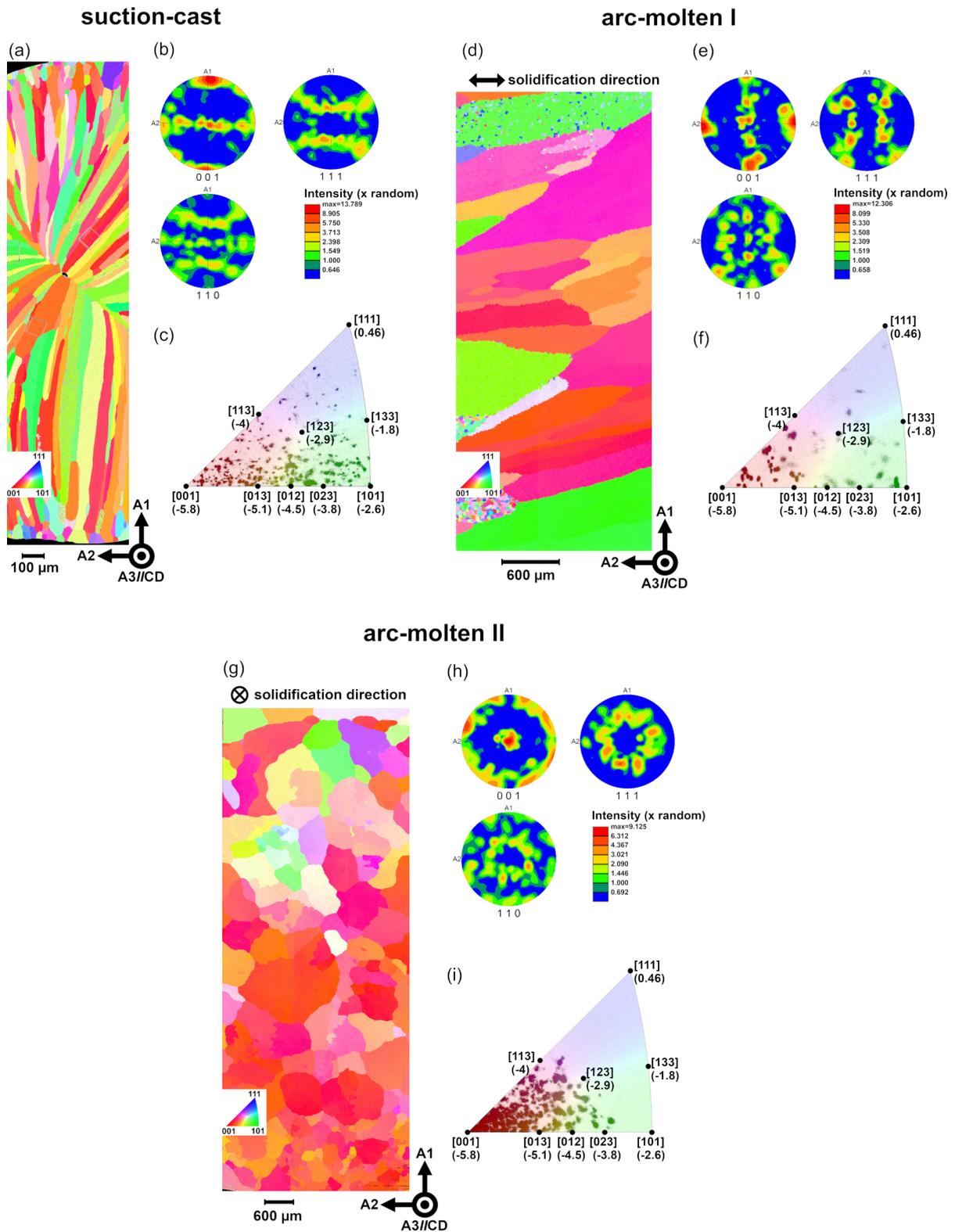

*Figure 2: (a), (d), (g) EBSD inverse pole figure maps of suction-cast and arc-molten Ni-Mn-In along the longitudinal plane (arc-molten I) and along the transversal plane (arc-molten II). (b), (e), (h) corresponding pole figure texture plots with logarithmic color scale and (c), (f), (i) inverse pole figures in compression direction (CD). In (c), (f), (i) the values in round brackets correspond to the theoretical stress-induced martensitic transformation strains in % for certain crystallographic directions.*

### 3.3. Mechanical properties

Figure 3 shows the compressive stress-strain curves of suction-cast, arc-molten I and arc-molten II Ni-Mn-In 10 K above the austenite finish temperature $A_f$. It should be emphasized that the specimens are of similar composition and transition temperature. Upon loading, the austenite deforms elastically until at a critical stress $\sigma^{Ms}$ the onset of the martensitic transformation is noticed. The stress-induced martensitic transformation exhibits a characteristic plateau. After completion of the transition at $\sigma^{Mf}$, initially elastic deformation and subsequently dislocation-mediated slip of the martensite can be observed which results in fracture at the maximum compressive stress $\sigma_{comp}$.

When the stress-strain curves are compared, a perfect agreement in the stress-induced martensitic transformation behavior of suction-cast and arc-molten I is apparent. This can be attributed to the similar texture in compression direction (CD) which has been shown in the previous section. It is worth mentioning that this coincidence is observed despite the significantly refined grain diameter after suction casting. By the double tangent method a transformation strain $\varepsilon_{tr}$ of 3.8 % can be determined for both samples. In arc-molten II, $\varepsilon_{tr}$ is increased to 4.4 % due to the enhanced <001> texture in CD. This results in a reduction of $\sigma^{Ms}$ from 174 MPa to 150 MPa as compared to suction-cast and arc-molten I. Besides that, a smaller transition slope is found in arc-molten II which can also be attributed to its texture [48]. As a consequence, $\sigma^{Mf}$ of 287 MPa in suction-cast and arc molten I is lowered to 247 MPa in arc-molten II.

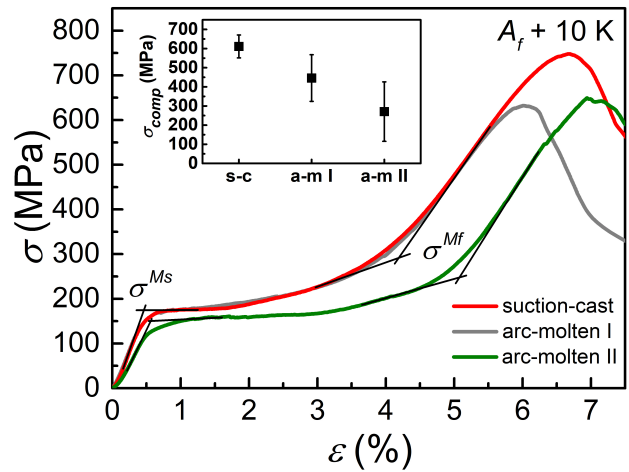

*Figure 3: Compressive stress-strain σ(ε) curves of suction-cast, arc-molten I and arc-molten II Ni-Mn-In of similar composition and transition temperature. The critical martensite start $\sigma^{Ms}$ and finish $\sigma^{Mf}$ stresses are indicated. The measurements were performed 10 K above $A_f$. The inset shows the average compressive stresses $\sigma_{comp}$ which have been obtained from at least three measurements of suction-cast, arc-molten I and arc-molten II.*

Besides the stress-induced martensitic transformation behavior, the compressive strength $\sigma_{comp}$ of the material is a crucial for its application in elastocaloric and multicaloric cooling to avoid structural fatigue during cyclic operation. While in Figure 3 the suction-cast specimen exhibits a $\sigma_{comp}$ of 748 MPa, it is reduced to 632 MPa in arc-molten I and to 649 MPa in arc-molten II. For better reliability, we measured at least three Ni-Mn-In specimens for each microstructure. It is worth mentioning, that for each microstructure the additional measurements exhibit a similar stress-induced martensitic transformation behavior as shown in Figure 3. The resulting average values of $\sigma_{comp}$ are illustrated in the inset of Figure 3. It is apparent that suction-cast specimens show the highest average value of $\sigma_{comp}$ (611 ± 60 MPa) which is attributed to the refined grain diameter. In arc-molten I, $\sigma_{comp}$ is reduced to 446 ± 122 MPa. A further decrease of $\sigma_{comp}$ to 270 ± 155 MPa is found for arc-molten II which could result from a preferential cracking of grain boundaries along the compressive direction [50]. Accordingly, some arc-molten II specimens exhibited failure already during the stress-induced transition transformation. The low $\sigma_{comp}$ of arc-molten II can hinder the application for elastocaloric and multicaloric cooling though slightly lowered stresses are required for the stress-induced martensitic transformation. The best combination of $\sigma_{comp}$ and stress-induced martensitic transformation behavior is obtained when suction-cast Ni-Mn-In is used. For that reason, the caloric response of suction-cast Ni-Mn-In is investigated in more detail in the following.

### 3.4. Calorimetric characterization

In Figure 4(a), DSC curves of suction-cast Ni-Mn-In are displayed before and after mechanical training. The mechanical training has been carried out by five superelastic cycles up to 287 MPa at 305 K prior to the elastocaloric measurements shown in section 3.5. Upon cooling, both samples exhibit a strong exothermic peak corresponding to the martensitic transformation, while the endothermic peak upon heating characterizes the reverse martensitic transformation. In addition, a shoulder peak is visible during the martensitic transformation resulting from the characteristic microstructure of the suction-cast material which will be discussed later in more detail (see section 3.6.2). A thermal hysteresis $\Delta T_{hys}$ = $(A_s+A_f)/2-(M_s+M_f)/2$ of 7.7 K before training and of 9.8 K after training can be determined from the transition temperatures listed in Table 2. The slightly enlarged thermal hysteresis in the trained sample can result from dislocation accumulation during cyclic loading [51].

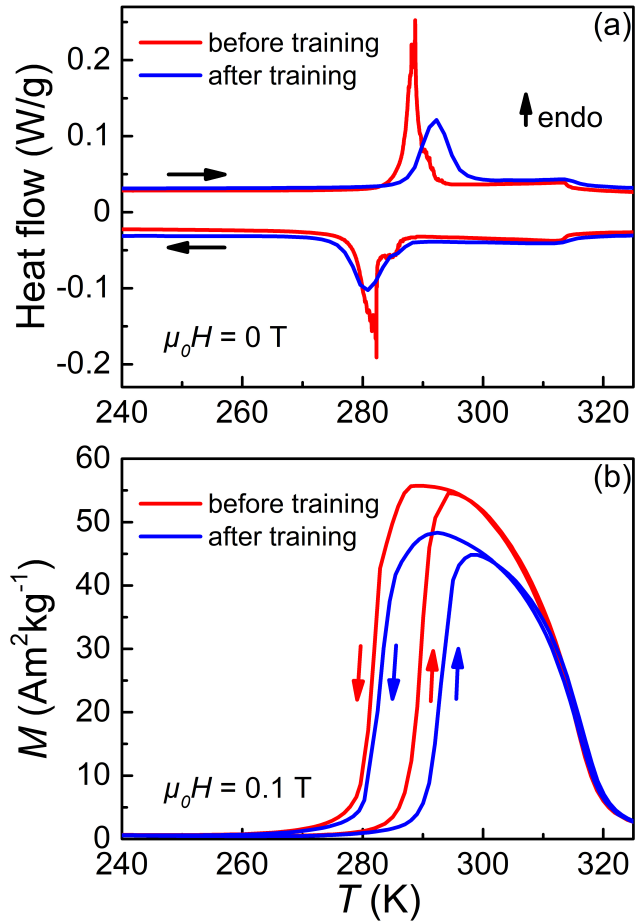

Figure 4: (a) DSC signals in zero field and (b) isofield curves of magnetization in a magnetic field of 0.1 T before and after mechanical training (by superelastic cycling) of suction-cast Ni-Mn-In.

Table 2: Transformation temperatures, i.e. $M_s$, $M_f$, $A_s$, $A_f$, and thermal hysteresis $\Delta T_{hys}$ of suction-cast Ni-Mn-In determined by DSC and M(T) curves in a field of 0.1 T before and after mechanical training. From the DSC measurements, the transition entropy changes $\Delta s_t$ were determined and the corresponding adiabatic temperature changes $\Delta T_{ad,DSC}$ at T = 305 K were estimated.

| Method | Transition temperatures | | | | $\Delta T_{hys}$ | $\Delta s_t$ cooling | $\Delta T_{ad,DSC}$ cooling | $\Delta s_t$ heating | $\Delta T_{ad,DSC}$ heating |
|---|---|---|---|---|---|---|---|---|---|
| | $M_s$ [K] | $M_f$ [K] | $A_s$ [K] | $A_f$ [K] | [K] | [Jkg$^{-1}$K$^{-1}$] | [K] | [Jkg$^{-1}$K$^{-1}$] | [K] |
| DSC before mechanical training | 286.7 | 277.8 | 285.6 | 294.2 | 7.7 | -19.5 ± 0.3 | +11.4 ± 0.2 | +21.2 ± 0.3 | -12.4 ± 0.2 |
| DSC after mechanical training | 287.9 | 276.4 | 287.6 | 296.3 | 9.8 | -18.4 ± 0.2 | +10.8 ± 0.1 | +20.2 ± 0.2 | +11.9 ± 0.1 |
| M(T) before mechanical training | 287 | 277.3 | 284.9 | 294.5 | 7.6 | | | | |
| M(T) after mechanical training | 288.3 | 277.9 | 288 | 296.7 | 9.3 | | | | |

The change of the calorimetric signal at 314 K is associated with the Curie temperature of the austenite $T_C^A$. The paramagnetic-ferromagnetic ordering around $T_C^A$ is more distinctly recognizable in the corresponding M(T) measurements in Figure 4(b) in a small field of 0.1 T, which also shows a good agreement with the martensitic transformation temperatures from DSC (see Table 2).

By integrating the calorimetric signal (baseline corrected heat flow per mass unit $\dot{Q}$) according to equation (1)

$$\Delta s_t = \int \frac{1}{T}(\dot{Q} - \dot{Q}_{baseline}) \left(\frac{dT}{dt}\right)^{-1} dT \qquad \text{Equation 1}$$

a transition entropy change $\Delta s_t$ of -19.5 ± 0.3 Jkg$^{-1}$K$^{-1}$ and 21.2 ± 0.3 Jkg$^{-1}$K$^{-1}$ is obtained in the non-trained sample for the martensitic and its reverse transformation, respectively. The difference in $|\Delta s_t|$ results from an increasing magnetic entropy contribution at lower temperatures and will be discussed later in more detail. In the trained sample $\Delta s_t$ is slightly lowered to -18.4 ± 0.2 Jkg$^{-1}$K$^{-1}$ and 20.2 ± 0.2 Jkg$^{-1}$K$^{-1}$ which can be attributed to the aforementioned accumulation of dislocations and/or the formation of retained martensite [51]. It should be emphasized that $\Delta s_t$ is a crucial parameter for the elasto- and magnetocaloric performance of the material. On the one hand, large values of $\Delta s_t$ enable high adiabatic temperature changes $\Delta T_{ad}$ upon the martensitic and its reverse transformation which can be estimated according to equation (2):

$$\Delta T_{ad,DSC} \simeq -\frac{T}{c_p}\Delta s_t \qquad \text{Equation 2}$$

with the temperature $T$ and the heat capacity $c_p$ (519.8 Jkg$^{-1}$K$^{-1}$) that has been extracted from our data on suction-cast Ni-Mn-In presented in [20]. For $T$ = 305 K (corresponds to temperature used for elastocaloric $\Delta T_{ad}$ measurements in section 3.5.1), a $\Delta T_{ad,DSC}$ of +11.4 ± 0.2 K and -12.4 ± 0.2 K before training and a $\Delta T_{ad,DSC}$ of +10.8 ± 0.1 K and -11.9 ± 0.1 K after training is assessed for the forward and reverse transformation, respectively (see Table 2). On the other hand, a rising $\Delta s_t$ results in an increasing slope $d\sigma^{Ms}/dT_t$ and therefore higher stresses to induce the martensitic transformation being described by the Clausius-Clapeyron equation

$$-\frac{\Delta s_t}{\varepsilon_{tr}} = \frac{1}{\rho}\frac{d\sigma^{Ms}}{dT_t} \qquad \text{Equation 3}$$

where $\rho$ is the mass density of the specimen. Note that equation (3) also illustrates the previously mentioned significance of the transformation strain $\varepsilon_{tr}$. Though $d\sigma^{Ms}/dT_t$ of the stress-induced martensitic transformation can be approximated by equation (3) (4.04 MPaK$^{-1}$ for $\Delta s_t$ = -19.5 Jkg$^{-1}$K$^{-1}$, $\varepsilon_{tr}$ = 3.8 %, and $\rho$ = 7.869 g/cm$^3$), more reliable and detailed information is obtained from isothermal $\sigma(\varepsilon)$ experiments at different temperatures (see section 3.5.1).

It should be emphasized that $\Delta s_t$ has a similar effect on the slope $\mu_0 dH/dT_t$ of the magnetic-field-induced reverse martensitic transformation, which is discussed in section 3.6.1.

### 3.5. Elastocaloric performance
#### 3.5.1. Isothermal response

In order to analyze the reversibility and the slope $d\sigma/dT_t$ ($d\sigma^{Ms}/dT_t$ and $d\sigma^{Mf}/dT_t$) of the stress-induced martensitic transformation in suction-cast Ni-Mn-In, isothermal loading/unloading measurements were performed at different temperatures as depicted in Figure 5(a). Note that the experiment was carried out after mechanical training of the sample by five superelastic cycles up to 287 MPa at 305 K. This explains the reduction of the critical stresses and the transformation strain $\varepsilon_{tr}$ compared to the values in Figure 3 and should ensure a repeatable material behavior to evaluate the effect of temperature on the stress-induced martensitic and its reverse transformation. Furthermore, a discontinuous temperature protocol, including heating to 310 K before setting the measurement temperature, was chosen to avoid the occurrence of minor loops [52].

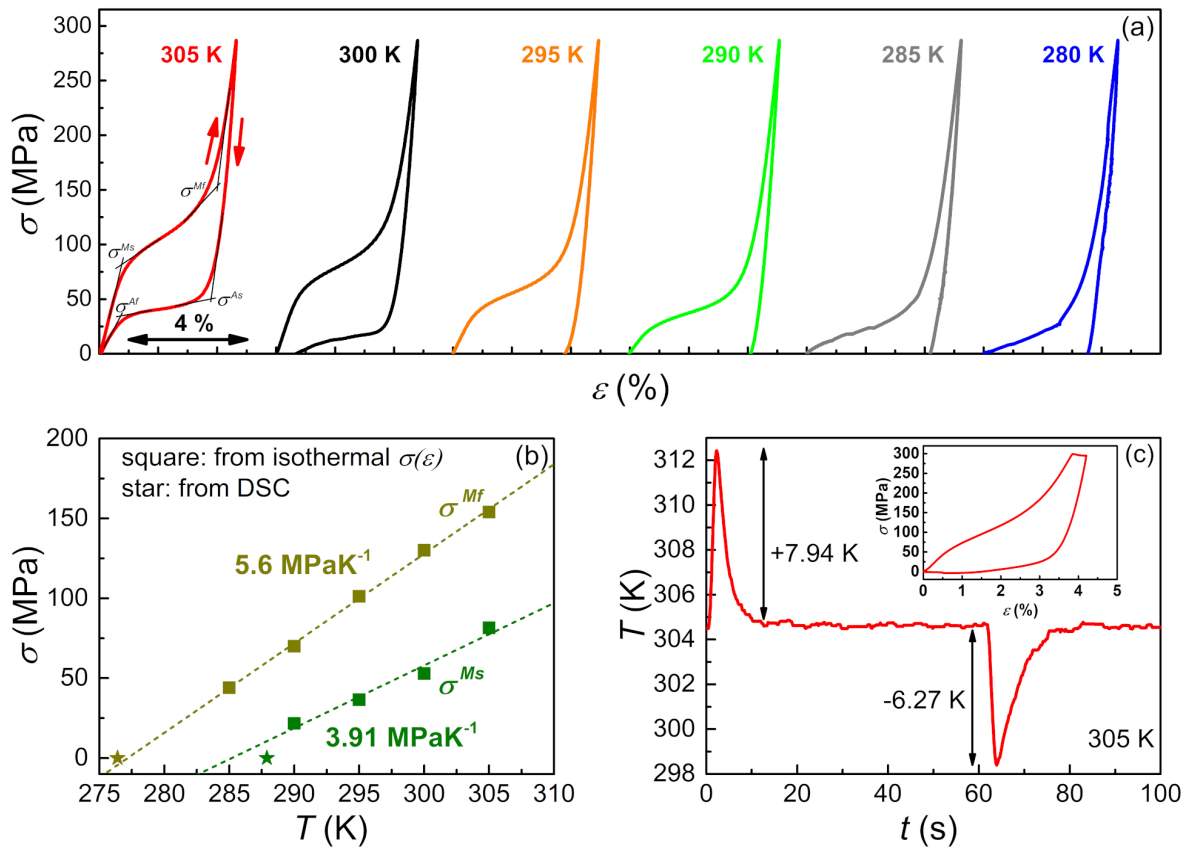

Figure 5: (a) Isothermal compressive stress-strain $\sigma(\varepsilon)$ response of suction-cast Ni-Mn-In up to 287 MPa at different temperatures. The critical martensite start $\sigma^{Ms}$, finish $\sigma^{Mf}$ and austenite start $\sigma^{As}$, finish $\sigma^{Af}$ stresses are indicated for the measurement at 305 K. (b) Corresponding temperature dependence of the critical martensite start $\sigma^{Ms}$ and finish $\sigma^{Mf}$ stresses. The data points at zero stress were taken from the DSC measurement after mechanical training. (c) Temperature-time T(t) profile upon loading and unloading under quasi-adiabatic conditions at a starting temperature of 305 K. In the inset, the corresponding stress-strain $\sigma(\varepsilon)$ curve is shown.

Figure 5(a) shows a consecutive reduction of $\sigma^{Ms}$ and $\sigma^{Mf}$ when the test temperature gets closer to $M_s$ = 287.9 K and $M_f$ = 276.4 K of the mechanically trained sample. The decrease in critical stresses can be attributed to a reduction of the relative stability of the austenite compared to the martensite phase at lower temperatures. In the corresponding critical stress-temperature phase diagram (Figure 5(b)), $\sigma^{Ms}$ exhibits a slope of 3.91 MPaK$^{-1}$, which is in good agreement with the value (4.04 MPaK$^{-1}$) obtained via the Clausius-Clapeyron equation (equation (3)). For $\sigma^{Mf}$ a much bigger slope of 5.6 MPaK$^{-1}$ can be observed. As a consequence, the stress-induced transition width upon isothermal mechanical loading is reduced for lower test temperatures. Accordingly, the temperature transition width upon isostress thermal cycling is reduced for lower stresses. When the test temperature is between $M_s$ and $M_f$, as for 285 K and 280 K, reorientation of thermally-induced martensite can occur upon loading [53,54]. As at 280 K the sample is predominantly martensitic before load application, the corresponding value of $\sigma^{Mf}$ might not be reliable and is therefore not included in the critical stress-temperature phase diagram.

Upon unloading, Figure 5(a) exhibits a full shape recovery at 305 K well above $A_s$ = 287.6 K and $A_f$ = 296.3 K at zero stress. When the test temperature is decreased to 300 K, the critical stress for the onset of the reverse transformation $\sigma^{As}$ is reduced due to an increasing stability of martensite at lower temperatures. Thereby, the reverse transformation is incomplete with an irrecoverable strain of 0.66 %. Between $A_f$ and $A_s$ at 295 K and 290 K no reverse transformation can be noticed. The recovery of the sample to its initial length upon heating to 310 K after the measurement implies no austenite slip resulting in irreversibility. A similar material response was reported for Ni-Fe-Ga single crystals and can result from the friction behavior of the shape-memory alloy [47]. Additionally, the reverse transformation is limited below $A_f$ and fully hindered below $A_s$ by the thermal hysteresis. Consequently, no shape recovery can be noticed at 285 K and 280 K. As a result of the irreversibilities at the majority of the test temperatures, no stress-temperature phase diagram can be determined for the reverse transformation.

For elastocaloric applications, the material should fully transform back to austenite upon unloading which means that a temperature well above $A_f$ has to be chosen. In this case, higher critical stresses and an increasing stress-induced transition width arise for the martensitic transformation compared to lower operation temperatures (see Figure 5(b)). These are utilized in multicaloric applications such as the "exploiting-hysteresis cycle" which works in the thermal hysteresis regime. If exemplarily $\sigma^{Mf}$ at 305 K and 285 K is compared, a decrease of about 70 % can be observed for the lower temperature. Furthermore, it should be emphasized that the irreversibility upon unloading at temperatures below $A_f$ as well as the slight increase in thermal hysteresis by mechanical training is desirable for the exploiting-hysteresis concept.

### 3.5.2. Adiabatic response

In order to evaluate the elastocaloric performance of suction-cast Ni-Mn-In, we performed mechanical cycling experiments under quasi-adiabatic conditions at 305 K. For this purpose, the specimen was loaded up to 300 MPa and unloaded with a strain rate of 3x10$^{-2}$ s$^{-1}$, which is two orders of magnitude higher than in the isothermal experiments. Additionally, a holding time of 60 s for thermal relaxation was applied. Figure 5(c) exhibits the corresponding temperature-time profile and stress-strain curve (inset). Upon loading, a $\Delta T_{ad}$ of 7.94 K can be noticed, while for the unloading segment $\Delta T_{ad}$ reduces to -6.27 K due to frictional heating [55]. The observed values are among the highest reported so far for Ni-Mn-In-(Co) compounds [24,56–58].

However, still a clear difference can be noticed when the obtained values are compared to the temperature changes $\Delta T_{ad,DSC}$ of the trained sample estimated by DSC curves (see Table 2). The origin of the difference in measured and estimated temperature change can result from the incomplete stress-induced transformation which is indicated by the continuation of the transition during the holding time (see inset of Figure 5(c)) and a lack of adiabaticity.

It should be pointed out that under isothermal conditions a stress of 154 MPa is enough to fully induce the martensitic transformation at 305 K, while under quasi-adiabatic conditions 300 MPa are not sufficient. The reason for this significant increase is the self-heating of the specimen under quasi-adiabatic conditions resulting in a rising transformation slope (and thus a rising stress-induced transition width) [59]. Accordingly, a self-cooling effect can be observed upon unloading. Overall, this results in a drastically increased stress hysteresis compared to the isothermal measurement shown in Figure 5(a) [60].

### 3.5.3. Isostress measurements

Complementary to the isothermal and quasi-adiabatic stress-strain measurements, isostress strain-temperature $\varepsilon(T)$ scans were performed up to 75 MPa with a fresh specimen of suction-cast Ni-Mn-In. The $\varepsilon(T)$ measurements at different constant loads are preferably used for the determination of isothermal entropy changes $\Delta s_T(T,\sigma)$ as possible overestimations can be excluded compared to the values assessed from isothermal stress-strain curves [61]. In this case, equation (4) is used for the computation of the isothermal entropy change:

$$\Delta s_T(T,\sigma) = \frac{1}{\rho} \int_0^\sigma \left(\frac{\partial \varepsilon(T,\sigma)}{\partial T}\right)_\sigma d\sigma. \quad \text{Equation 4}$$

In addition, the $\varepsilon(T)$ scans allow the determination of the stress-dependent critical temperatures for the reverse transformation which were not obtained from the stress-strain curves due to the aforementioned irreversibilities at test temperatures below 305 K.

Figure 6(a) shows the $\varepsilon(T)$ measurements at constant compressive stresses of 1, 12.5, 25, 50, and 75 MPa. The applied stress was increased from low to high values in consecutive thermal cycles (cooling from austenite to martensite and subsequent heating from martensite to austenite). It is apparent, that with a higher stress the transformation strain $\varepsilon_{tr}$ rises which originates from the increasing formation of favorably oriented martensite variants. Thereby, the transformation exhibits a completely recoverable behavior up to 50 MPa, while upon cycling at 75 MPa a permanent strain of 0.75 % can be seen. In addition, Figure 6(a) displays a significant increase in transition temperatures and thermal hysteresis when the stress is raised. The shift of the transformation ($dT_t/d\sigma$) towards higher temperatures originates from the stabilization of martensite by uniaxial load due its lower symmetry and volume. The enlargement of the thermal hysteresis can be ascribed to an increasing frictional work and elastic strain energy dissipation as well as dislocation formation during the martensitic transformation [62]. As a consequence, higher temperatures are required to promote the reverse transformation. This behavior is depicted in the temperature-stress phase diagram (see Figure 6(b)) by the significantly enhanced linear shift of the reverse transformation (0.3 KMPa$^{-1}$) compared to the forward transformation (0.21 KMPa$^{-1}$) with applied stress. Thus, the thermal hysteresis increases by 0.09 KMPa$^{-1}$. It should be emphasized, that the above described energy-dissipation mechanisms counteract a declining thermal hysteresis which is expected at higher temperatures. According to equation (3), higher transition entropy changes $\Delta s_t$ should result in a smaller shift of the transition temperature with applied stress ($dT_t/d\sigma$). In consequence of its lower $\Delta s_t$ (see Table 2), a higher shift of the forward martensitic transformation than of its reverse transformation can be expected, which should reduce the thermal hysteresis upon increasing stress. The inset of Figure 6(b) provides the stress-dependent transition temperatures of $M_s$, $M_f$, $A_s$ and $A_f$.

The change of $M_s$ (0.27 KMPa$^{-1}$) and $M_f$ (0.15 KMPa$^{-1}$) are in good agreement with the data obtained from isothermal measurements (see Figure 5(b)). Accordingly, the isostress curves also display an increasing temperature-induced transition width at higher stresses for the martensitic transformation. A similar behavior can be observed for the reverse transformation which exhibits a shift of $A_s$ by 0.25 KMPa$^{-1}$ and $A_f$ by 0.37 KMPa$^{-1}$. A possible explanation of the transition broadening with higher loads are the aforementioned energy dissipation mechanisms and/or the increasing formation of dislocations and stress inhomogeneities at defects [62].

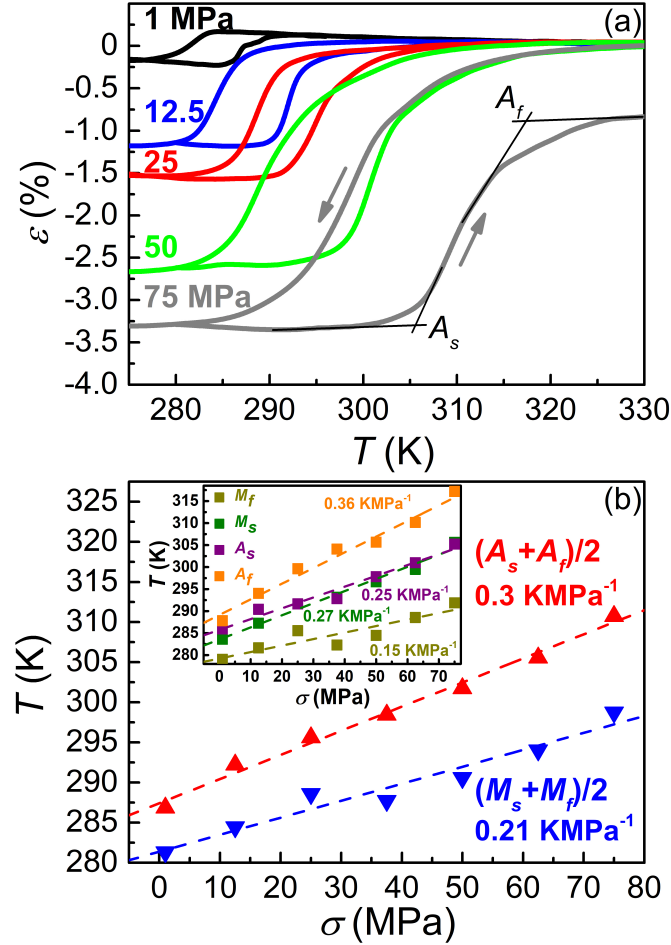

Figure 6: (a) Isostress strain-temperature ε(T) curves of suction-cast Ni-Mn-In at different compressive stresses. The determination of the reverse martensitic transformation temperatures $A_s$, $A_f$ is exemplarily shown for the thermal cycle at 75 MPa. (b) Corresponding temperature-stress T(σ) phase diagram. The inset of (b) gives a more detailed depiction of the phase diagram including the respective martensitic transformation temperatures $M_s$, $M_f$, $A_s$, $A_f$ as a function of applied compressive stress.

The increase of both, thermal hysteresis and transition width, was observed for different polycrystalline Ni-Mn-based Heusler compounds subjected to uniaxial loads [14,56,63]. In contrast to that, these alloys show a declining thermal hysteresis and smaller transition widths for the application of hydrostatic pressure, being caused by an improved geometric compatibility [64]. Besides that, a significant difference in the shift of the transition temperature can be observed, depending whether uniaxial load or a hydrostatic pressure is applied. In case of Ni-Mn-In, the shift of the transition with uniaxial load is approximately one order of magnitude higher than for the application hydrostatic pressure [65]. The difference is caused by structural distortions upon the martensitic transformation which are dominated by shear when the material is subjected to uniaxial stress, but purely volumetric for hydrostatic pressure [66].

Figure 7 displays the isothermal entropy change $\Delta s_T(T,\sigma)$ of the forward transformation determined via equation (4). The isothermal entropy change of the reverse transformation could not be computed in a proper form due to a limited number of data points. Figure 7 exhibits that a change in stress of 75 MPa results in an isothermal entropy change of -11 Jkg$^{-1}$K$^{-1}$ corresponding to 56 % of the transition entropy change $\Delta s_t$ (see Table 2). Upon cyclic application a slight reduction of $|\Delta s_T(T,\sigma)|$ is expected due to the irrecoverable strain evolving during the martensitic transformation at 75 MPa. Nevertheless, significantly higher values are achieved in comparison with the application of 75 MPa hydrostatic pressure for which approximately 17 % of $\Delta s_t$ were reached in a Ni-Mn-In specimen of similar transition temperature [65].

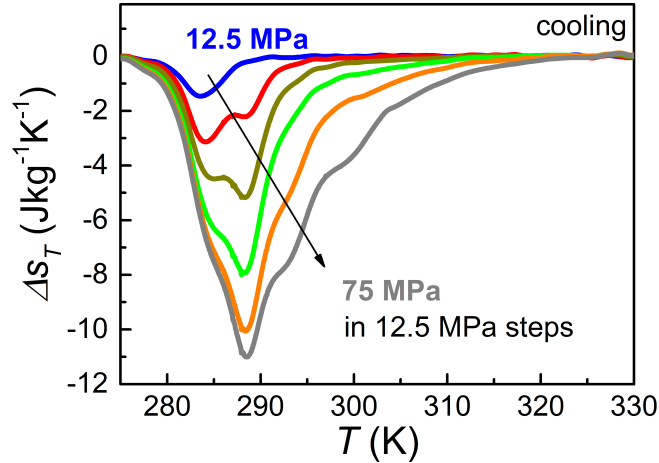

Figure 7: Temperature-dependent isothermal entropy change $\Delta s_T(T,\sigma)$ of martensitic transformation (upon cooling) for different applied compressive stresses of suction-cast Ni-Mn-In.

### 3.6. Magnetocaloric performance
#### 3.6.1. Indirect measurements

The magnetocaloric response of suction-cast Ni-Mn-In was checked by isofield, isothermal and adiabatic measurements. Note that the measurements were performed in a mechanically non-trained state of the material. In Figure 8(a), isofield $M(T)$ curves are exemplarily shown for 0.1, 2, 5 and 10 T. It is apparent that the first-order transformation shifts towards lower temperatures at higher magnetic fields, as the ferromagnetic austenite phase is stabilized. The corresponding temperature-magnetic field phase diagram in Figure 8(b) exhibits that the commonly used linear approximation for the change of transition temperature $T_t$ with magnetic field only holds true for small fields, but clearly deviates in high fields. Thereby, a slightly more pronounced shift $(dT_t/\mu_0 dH)$ can be noticed for the martensitic than for the reverse transformation resulting in a non-linear increase of the thermal hysteresis (see inset Figure 8(b)). In addition, the first-order transformation displays a broadening upon increasing fields and a distinct kink in magnetization. The kink is caused by the characteristic microstructure of the material and will be described later in more detail (see section 3.6.2).

The non-linearity in the change of $T_t$, the enhanced shift $(dT_t/\mu_0 dH)$ of the forward transformation and the broadening of the transition with magnetic field originate from a rising magnetic entropy change $\Delta s_{mag}$ at lower temperatures [67]. For the inverse magnetocaloric effect, $\Delta s_{mag}$ counteracts the lattice entropy change $\Delta s_{lat}$ which is approximated to be constant in the studied temperature range [68]. The electronic entropy change $\Delta s_{el}$ is negligible for Heusler alloys and thus not considered here [69]. As a result, the transition entropy change $\Delta s_t$ increasingly dilutes when the transition temperature is lowered by magnetic field or small compositional changes. According to Clausius-Clapeyron for magnetic-field-induced transformations (see equation (5)), the decrease in $\Delta s_t$ results in a rising $(dT_t/\mu_0 dH)$:

$$\frac{dT_t}{\mu_0 dH} = \frac{\Delta M(T)}{\Delta s_t(T)} = \frac{\Delta M(T)}{|\Delta s_{lat}| - |\Delta s_{mag}(T)|}. \quad \text{Equation 5}$$

The temperature dependence of $\Delta s_t$ can also be observed for the forward and the reverse transformation which exhibit a thermal hysteresis of 7.7 K at 0 T in the studied sample. Upon cooling (forward transformation) a $|\Delta s_t|$ of 19.5 Jkg$^{-1}$K$^{-1}$ and upon heating a $|\Delta s_t|$ of 21.2 Jkg$^{-1}$K$^{-1}$ was measured via DSC (see Table 2). This explains why the forward transformation shows a larger shift and the thermal hysteresis increases for rising magnetic fields. For the similar reason, a broadening of the transition in higher magnetic fields can be observed. In zero field, the transition temperature is spread over about 9 K. According to the temperature dependence of $(dT_t/\mu_0 dH)$, the parts of the sample transforming at lower temperatures exhibit a more pronounced shift than the parts of sample transforming at higher temperatures which results in a broadening of the transition in larger fields.

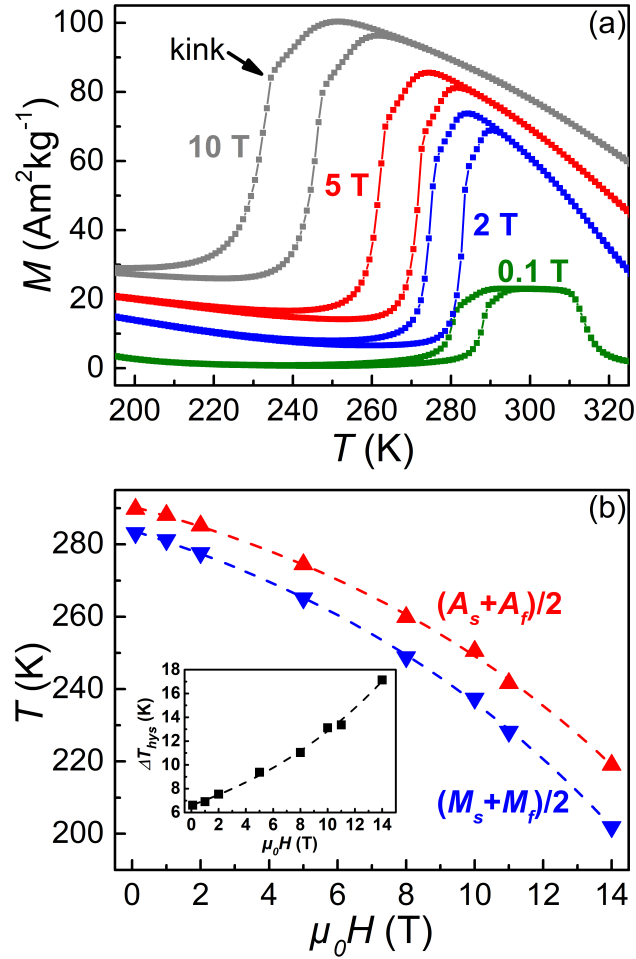

Figure 8: (a) Isofield curves of magnetization M(T) in various magnetic fields of suction-cast Ni-Mn-In. (b) Corresponding temperature-field T(H) phase diagram. The inset of (b) shows the field-dependence of the thermal hysteresis $\Delta T_{hys}(\mu_0 H)$.

Figure 9(a) illustrates isothermal M(H) curves up to 10 T of the magnetic-field-induced reverse martensitic transformation at various temperatures. Figure 9(b) exhibits the corresponding isothermal entropy change $\Delta s_T(T,H)$ which has been determined from the M(H) measurements up to 10 T and M(T) curves up to 2 T according to equation (6):

$$\Delta s_T(T,H) = \mu_0 \int_0^H \left(\frac{\partial M(T,H)}{\partial T}\right)_H dH. \quad \text{Equation 6}$$

The computed $\Delta s_T(T,H)$ of both methods exhibits a good agreement. For the maximum $\Delta s_T(T,H)$, a value of 19.6 Jkg$^{-1}$K$^{-1}$ can be found at 283 K for a field change of 4 T which corresponds to a complete phase transformation. A reasonable coincidence with $\Delta s_t$ from DSC (see Table 2) can be noticed. At 283 K just below $A_s$ = 285.6 K, field changes of less than 4 T are not sufficient to achieve a full transformation. At higher field changes, the transformation can be fully induced, but the material additionally shows an increasing conventional magnetocaloric effect counteracting the inverse effect. As a consequence, $\Delta s_T(T,H)$ decreases at field changes smaller and larger than 4 T. At temperatures above 283 K the decay of $\Delta s_T(T,H)$ can be attributed to a decrease of the transformed phase fraction resulting from the formation of temperature-induced austenite before field application. In addition, the conventional magnetocaloric effect of the austenite increases at higher temperatures as the Curie temperature $T_C^A$ = 314 K is approached. This effect is indicated by the strongly negative values of $\Delta s_T(T,H)$ at high temperatures in Figure 9(b) and further decreases $\Delta s_T(T,H)$ of the reverse martensitic transformation. At temperatures below 283 K, the already discussed decrease of $\Delta s_t$ causes the declining $\Delta s_T(T,H)$.

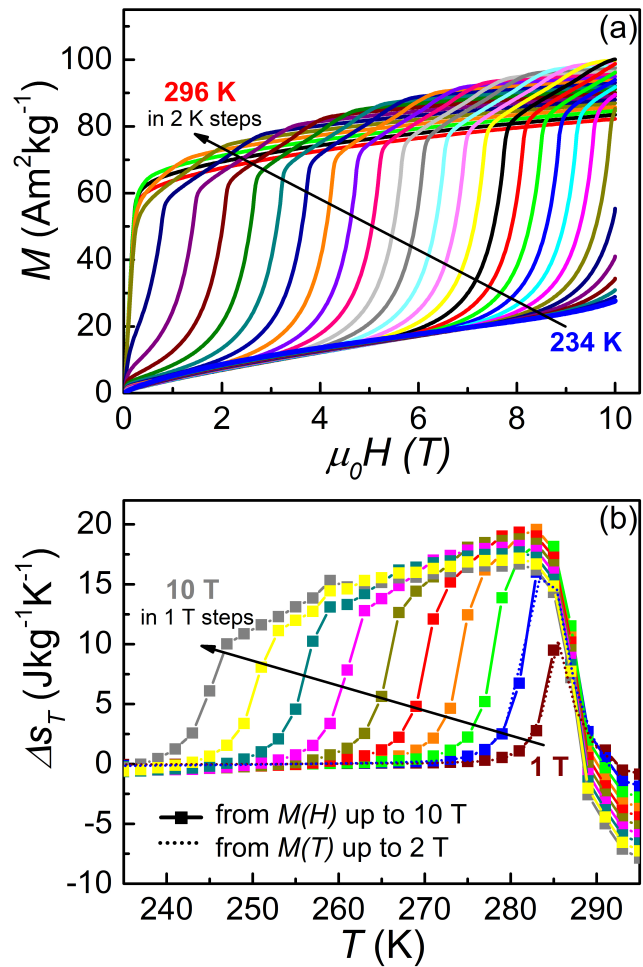

*Figure 9: (a) Isothermal measurements of magnetization at different temperatures of suction-cast Ni-Mn-In. (b) Temperature-dependent isothermal entropy change $\Delta s_T(T,H)$ of reverse martensitic transformation for different applied magnetic fields.*

### 3.6.2. Two-step transformation behavior

In order to investigate the aforementioned distinct kink in magnetization of the first-order transformation which can be seen in the *M(T)* and *M(H)* measurements, we performed temperature-dependent optical microscopy. Figure 10 shows the microstructural evolution of the suction-cast cylinder upon cooling. At 290 K, the sample is fully in the austenitic state. When the temperature is lowered, the fine equiaxed grains in the chill zone, at the edge of the sample, start to transform first and form a shell that surrounds the austenite in the sample center. In the *M(T)* curve of the full cylinder, this part corresponds to the kink at the onset of the transformation between 290 K and 282 K (see Figure 11(a)). The *M(T)* curve of the pure chill zone (extracted by cutting) in Figure 11(b) shows that its transformation is mostly finished at 282 K. According to Figure 11(a), the chill zone contributes 33 % of the samples magnetization which is in good agreement with the transformed phase fraction obtained from the micrograph at 282 K.

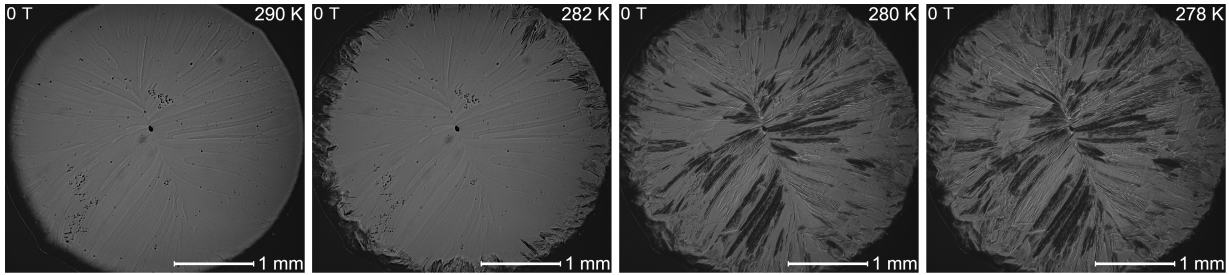

*Figure 10: Temperature-dependent evolution of the microstructure of suction-cast Ni-Mn-In upon cooling in zero field.*

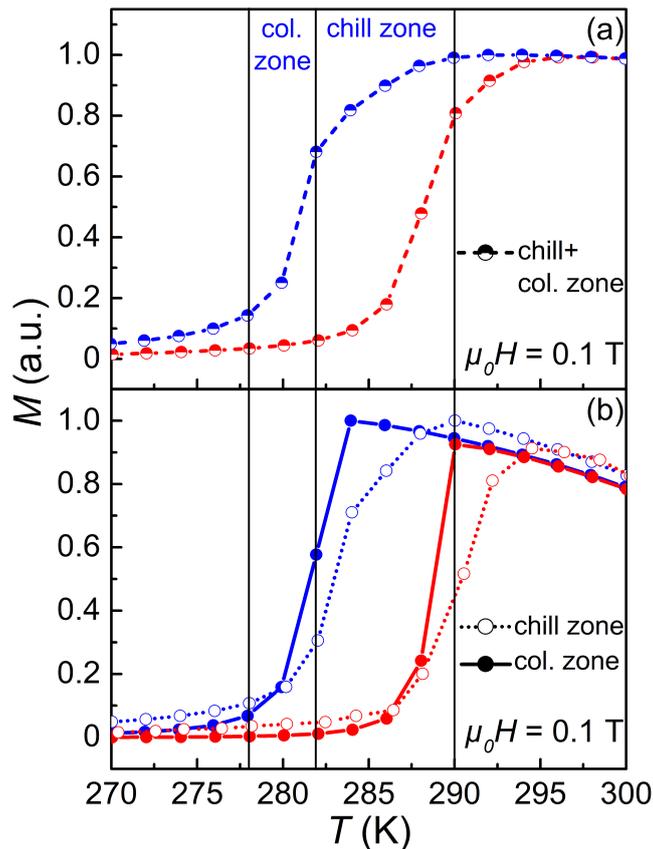

*Figure 11: Isofield curves of magnetization M(T) in a magnetic field of 0.1 T of (a) the complete suction-cast cylinder, (b) the pure chill zone (edge) and pure columnar (col.) zone (center) of the cylinder, respectively. The separate pieces of chill and columnar zone were extracted by cutting. Blue symbols and lines correspond to cooling and red symbols lines, to heating.*

Upon further cooling, a highly directional transformation process along the columnar grains can be observed (see supplementary video). Thereby, the parent austenite transforms within only 4 K. This can also be seen in the *M(T)* curve of the extracted columnar zone (see Figure 11(b)) with a slight difference in absolute temperature compared to the optical microscopy. Upon heating, the reverse transformation behavior can be noticed. In this case, the martensite in the columnar zone of the sample is shrinking first and subsequently the martensite of the chill zone. The reason for the two-stage transformation is a small difference in the chemical composition of the columnar and the chill zone which was detected by an EDX linescan (not shown here). While the columnar zone is chemically homogeneous, the chill zone shows a gradual increase of about 0.1 at% Ni and 0.15 at% In as well as a decrease of ca. 0.25 at% Mn which explains the broadened transformation of the chill zone. In addition, the significantly smaller grain size in the chill zone of the sample can contribute to the difference in transition temperature and width [70]. This means that even smaller critical magnetic fields can be realized to fully induce the reverse transformation when the chill zone is removed by a simple cylindrical grinding procedure.

### 3.6.3. Direct measurements

The complete evaluation of the compound's magnetocaloric performance requires information on the field-induced adiabatic temperature change $\Delta T_{ad}$ besides the $\Delta s_T(T,H)$ data. For this purpose, we carried out $\Delta T_{ad}$ measurements in a setup of two nested Halbach magnets with a maximum field change of 1.9 T and in pulsed magnetic fields up to 10 T. While the $\Delta T_{ad}$ measurements in the Halbach magnet provide an estimation of the materials applicability for conventional magnetic refrigeration, the pulsed-field data is used to check its suitability for the exploiting-hysteresis concept. The latter enables higher fields and faster sweeping rates [21]. Simultaneously with $\Delta T_{ad}$ the transformation strain $\varepsilon_{tr}$ was recorded which allows studying kinetical effects of the field-induced reverse martensitic transformation [45].

#### 3.6.3.1. Field dependence of the adiabatic temperature change

Figure 12(a) shows the field dependence of $\Delta T_{ad}(H)$ measured in the Halbach setup (symbols) and in pulsed fields (lines) of 2, 5, and 10 T for an initial sample temperature $T_{start}$ of 285 K. Note that at this temperature the sample is close to $A_s$ = 285.6 K. At first, the materials response to the 10 T pulse is discussed. Upon field application, the onset of the reverse martensitic transformation is indicated by a rising $|\Delta T_{ad}|$ at the critical field $H^{As}$. A $\Delta T_{ad}$ of -7.3 K can be observed when the transition is completed at a $H^{Af}$ of about 7.5 T. Above $H^{Af}$, the sample temperature rises when the magnetic field is increased further due to the conventional magnetocaloric effect around $T_C^A$. The same effect results in a linear cooling down to -9.3 K upon field removal until the transformation back to martensite takes place at a $H^{Ms}$ of about 1 T. This is equivalent to an increase of $|\Delta T_{ad}|$ by 27 %, which can be utilized in the exploiting-hysteresis cycle. For that purpose, the martensitic transformation has to be fully hindered by the thermal hysteresis. In the present sample, the thermal hysteresis causes a partial martensitic transformation with a $\Delta T_{ad}$ of -4.9 K when the magnetic field is fully removed. The 5 T pulse also exhibits a completed transformation from martensite to austenite when the magnetic field is applied. This is indicated by the coincidence with the signal of the 10 T pulse upon field removal. The counteracting inverse and conventional magnetocaloric effect results in a different $\Delta T_{ad}$ at the maximum field of 5 T (-8.2 K) and 10 T pulse (-6.7 K) which is in agreement with the observations for $\Delta s_T(T,H)$ (see Figure 9). In case of the 2 T pulse and the measurement in the Halbach setup (1.9 T), the field changes are not sufficient to fully induce a reverse martensitic transformation. Hence, a distinct minor-loop behavior can be noticed. The achieved $\Delta T_{ad}$ values for the different field changes and $T_{start}$ are summarized in Figure 13 and will be discussed in detail later.

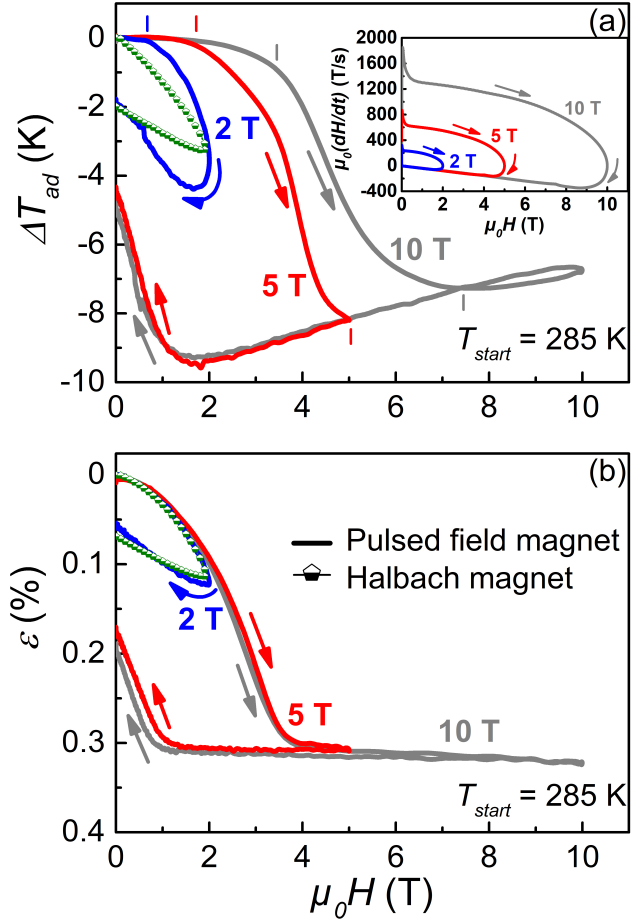

*Figure 12: (a) Adiabatic temperature change $\Delta T_{ad}$ of suction-cast Ni-Mn-In in a magnetic field of 1.9 T (green symbols, measured in Halbach setup) and in pulsed magnetic fields of 2, 5, and 10 T (solid lines) at $T_{start}$ = 285 K. The bars indicate onset $H^{As}$ and finish $H^{Af}$ of the field-induced reverse martensitic transformation in the pulsed-field measurements. The inset displays the pulsed-field sweep-rates $\mu_0(dH/dt)$ depending on the magnetic field. (b) Corresponding strain $\varepsilon$ as a function of the different magnetic fields.*

### 3.6.3.2. Magnetic-field-induced transformation dynamics

For the pulsed-field experiments, the maximum magnetic field was always reached after 13 ms corresponding to maximum sweeping rates of 1850, 866, and 318 Ts$^{-1}$ for 10, 5, and 2 T, respectively (see inset of Figure 12(a)). On the contrary, the field removal is significantly slower and the sweeping rates coincide. In the Halbach setup, the maximum magnetic field of 1.9 T is applied and removed with a considerably smaller rate of 1 Ts$^{-1}$. The large span of field-sweep rates from 1 Ts$^{-1}$ up to 1850 Ts$^{-1}$ during field application allows to study dynamical effects of the field-induced reverse martensitic transformation. For that purpose, the field has to be applied at the same initial sample temperature $T_{start}$. Figure 12(a) displays for $T_{start}$ = 285 K a growing $H^{As}$ upon increasing field-sweep rate. The same holds true for $H^{Af}$ if the reverse martensitic transformation is completed (see 5 T and 10 T pulses). However, by $\Delta T_{ad}$ measurements at $T_C^A$ (not shown here) a delay of the thermocouple could be revealed for the pulsed-field measurements, which can be attributed to the thermal mass and conductance of the thermocouple [71]. This causes an increasing field hysteresis with rising field-sweep rates. As a consequence, the $\Delta T_{ad}$ data do not allow a profound analysis of the transformation dynamics. Instead, the simultaneously detected transition strains in Figure 12(b) need to be considered. The instantaneous strain response upon the transformation confirms that the increase of $H^{As}$ and $H^{Af}$ upon rising field-sweep rates can be attributed to the thermocouple delay and not to dynamical effects of the transformation. Thus, an immediate onset of the reverse martensitic transformation can be observed in Figure 12(b) for all sweeping rates upon field application which agrees with the "delay-free" $\Delta T_{ad}$ measurement in the Halbach setup. In addition, $H^{Af}$ coincides for maximum field rates of 866 Ts$^{-1}$ (5 T pulse) and 1850 Ts$^{-1}$ (10 T pulse). It should be emphasized, that the coincidence in $H^{Af}$ is significantly different from the behavior observed in conventionally arc-molten Ni-Mn-In, where $H^{Af}$ increases at field rates higher than 865 Ts$^{-1}$ [45]. This may be explained by the higher defect density of the fine-grained suction-cast compared to the coarse-grained arc-molten microstructure which facilitates the annihilation of the martensite during the field-induced reverse martensitic transformation. Hence, no dynamical effects of $H^{Af}$ can be seen in the studied field-sweep range. Note that the strain measurement of the suction-cast material was performed in the cutting surface of the suction-cast cylinder which is perpendicular to the one shown in Figure 10. Further strain measurements in the cross section are planned to investigate the influence of the anisotropic transformation behavior observed by temperature-dependent microscopy (see Figure 10). Besides that, no two-step transformation behavior can be seen by a distinct kink in the detected $\Delta T_{ad}$ and strain signal, because both quantities were measured in the columnar zone of the sample.

### 3.6.3.3. Maximum and cyclic adiabatic temperature change

Figure 13(a) shows exemplarily a field-dependent $\Delta T_{ad}$ measurement at $T_{start}$ = 281 K in the Halbach setup to introduce the different quantities of $\Delta T_{ad}$ which are discussed in the following. In order to investigate the maximum $\Delta T_{ad}$ of the field-induced reverse martensitic transformation, $\Delta T_{ad}$ was determined at the maximum field $H_{max}$ upon the first field application. An exception was made for incomplete phase transitions at which the maximum value of $|\Delta T_{ad}|$ occurs at a field slightly different from $H_{max}$ (see e.g. 2 T pulsed-field measurement in Figure 12(a)). In this case the maximum $|\Delta T_{ad}|$ was taken. The values of $\Delta T_{ad}$ upon the first field application are summarized in Figure 13(b) as a function of the initial sample temperature for field changes of 1.9, 2, 5, and 10 T. The grey area exhibits the temperature region of the reverse martensitic transformation in zero field which was obtained via DSC (see Table 2). The highest $\Delta T_{ad}$ of -9.2 K can be observed at $T_{start}$ = 280 K for a field change of 5 T corresponding to a full reverse martensitic transformation. Compared to the estimated $\Delta T_{ad}$ of -11.4 K from DSC (for $T$=280 K), the measured value is slightly lower. A possible explanation for the difference is an increased $\Delta s_{mag}$ at $T_{start}$ = 280 K which reduces $\Delta s_t$ compared to the thermally driven transformation in DSC taking place at higher temperatures. In addition, the combination of inverse and conventional magnetocaloric effect in the measured sample results in a $\Delta T_{ad}$ falling below the predicted value. If $\Delta T_{ad}$ and the required field change are compared to arc-molten Ni-Mn-In of a similar transition temperature a good agreement is found [45]. For the dependence of $\Delta T_{ad}$ on $T_{start}$ and the applied magnetic field, similar trends as for $\Delta s_T(T,H)$ in Figure 9 can be noticed. Thus, at $T_{start}$ lower than 280 K $|\Delta T_{ad}|$ decays due to a gradual decrease of $\Delta s_t$ and the transformed phase fraction. At temperatures higher than 280 K, the reduction of $|\Delta T_{ad}|$ can be attributed to an enhanced conventional magnetocaloric effect and also to a decrease of the transformed phase fraction. The latter results from the formation of temperature-induced austenite in the sample before field application. The lower $\Delta T_{ad}$ (-8.2 K) compared to $T_{start}$ = 280 K and the immediate onset of the transformation (see Figure 12(b)) indicate also the presence of some temperature-induced austenite at $T_{start}$ = 285 K. For the $\Delta T_{ad}$ in the Halbach setup with field changes of 1.9 T and the 2 T pulses a good agreement can be noticed. The partial transformation exhibits a maximum $\Delta T_{ad}$ of -4.4 K in the Halbach and -4.7 K in the 2 T pulse at a $T_{start}$ of 287 K.

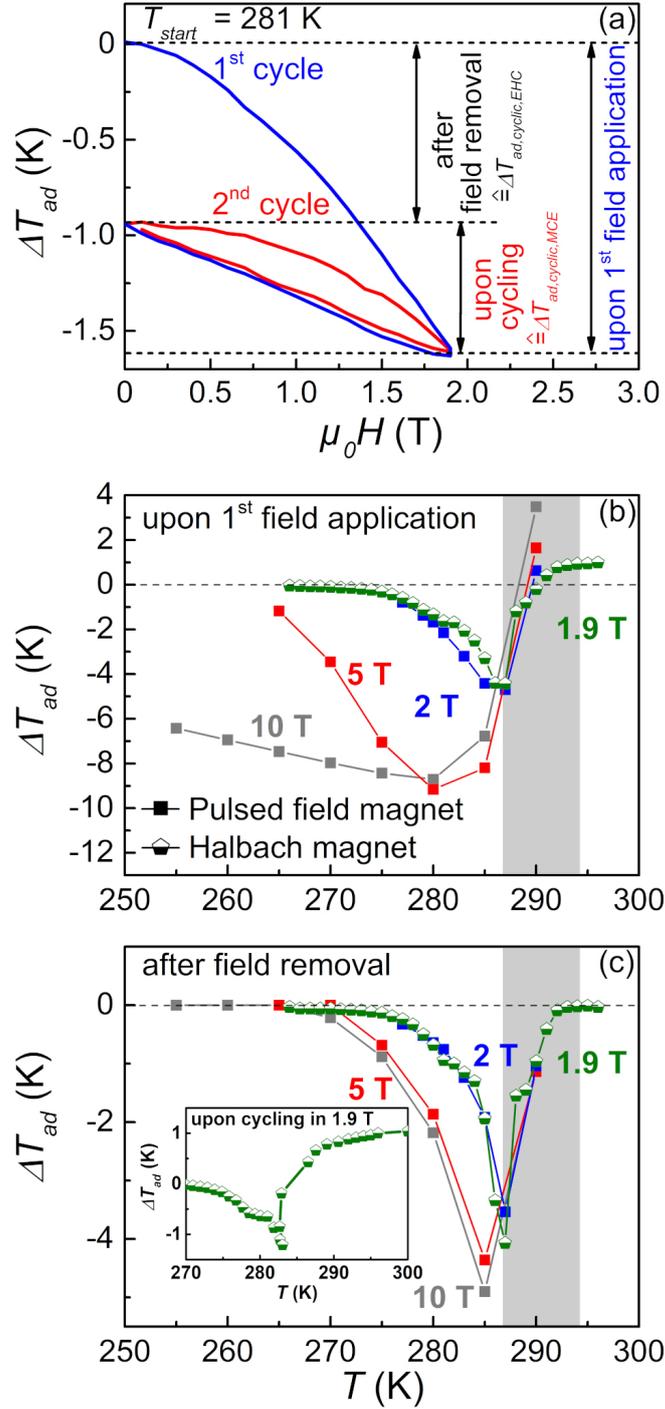

*Figure 13: (a) Field dependence of the adiabatic temperature change $\Delta T_{ad}$ in suction-cast Ni-Mn-In for an initial sample temperature $T_{start}$ of 281 K. The measurement was performed in a Halbach setup and visualizes the difference in $\Delta T_{ad}$ between the 1$^{st}$ and 2$^{nd}$ cycle (field application and removal). (b) Adiabatic temperature change $\Delta T_{ad}$ upon the 1$^{st}$ field application as a function of the initial sample temperature for maximum field changes of 1.9 T (green symbols, measured in a Halbach setup), 2, 5, and 10 T (solid lines and squares, measured in pulsed-field magnet). (c) Corresponding adiabatic temperature change $\Delta T_{ad}$ which remains after the magnetic field is fully removed. The inset shows the adiabatic temperature change $\Delta T_{ad}$ upon cycling as a function of the initial sample temperature for a field change of 1.9 T. The grey area illustrates the temperature region of the reverse martensitic transformation.*

In order to check the applicability of suction-cast Ni-Mn-In for conventional magnetocaloric cooling, $\Delta T_{ad}$ was determined upon cycling in 1.9 T, the field change that can be provided by permanent magnets. For that purpose, the 2$^{nd}$ cycle (field application and removal) in the Halbach magnet was analyzed as shown in Figure 13(a). It should be emphasized, that the initial sample temperature for the 1[st] and 2[nd] cycle can differ from each other when the 1[st] cycle is not fully reversible. The $\Delta T_{ad}$ values upon cycling are depicted in the inset of Figure 13(c) as a function of the initial sample temperature (for the 2[nd] cycles). A maximum cyclic effect of -1.2 K can be observed at 283 K which is in good agreement with the value reported for arc molten Ni-Mn-In [36]. However, the $\Delta T_{ad}$ upon cycling is significantly lower than the maximum effect during the first field application. The difference results from the thermal hysteresis which hinders a complete back transformation to martensite after field removal.

### 3.7. Exploiting-Hysteresis Cycle

The exploiting-hysteresis cycle benefits from the limited back transformation to martensite by utilizing the adiabatic temperature change $\Delta T_{ad}$ of the material after magnetic field removal (see Figure 13(a)). After the magnetic field is completely taken off, the heat is absorbed from the cooling compartment. In the next step, the material is transformed to its initial martensite state by the application of an uniaxial load before the cycle starts again with the application of the magnetic field. It should be emphasized, that the exploiting-hysteresis concept utilizes the adiabatic temperature change upon the field-induced reverse martensitic transformation (after field removal) while the stress-induced martensitic transformation only sets the material back to its initial martensite state without the necessity of achieving a $\Delta T_{ad}$. Hence, we can estimate the cyclic $\Delta T_{ad}$ which can be obtained by the exploiting-hysteresis concept from the $\Delta T_{ad}$ after field removal and the stress required to set the material back to its initial state by the critical stress-temperature phase diagram.

Figure 13(c) shows $\Delta T_{ad}$ after field removal as a function of the initial sample temperature for maximum fields of 1.9, 2, 5, and 10 T. For a field change of 1.9 T a maximum $\Delta T_{ad}$ of -4.1 K can be observed at $T_{start}$ = 287 K which is close to $\Delta T_{ad}$ (-4.4 K) upon the first field application. As the sample temperature coincides after the heat transfer with $T_{start}$ of 287 K, according Figure 5(b), a stress of 55 MPa has to be applied to bring the sample back to its original state. Hence, the $\Delta T_{ad}$ after field removal corresponds to the cyclic adiabatic temperature $\Delta T_{ad,cyclic,EHC}$ which can be utilized in the exploiting-hysteresis cycle (see also Figure 13(a)). $\Delta T_{ad,cyclic,EHC}$ can be increased further at higher magnetic fields. It should be emphasized that in comparison with conventional magnetic refrigeration the volume of permanent magnets can be reduced in the exploiting-hysteresis concept. In consequence, higher, as more focused, magnetic fields can be utilized which enable to increase the transformed phase fraction. For field changes of 5 and 10 T at a $T_{start}$ of 285 K, cyclic effects of -4.4 K and -4.9 K can be achieved, respectively. In both cases a stress of 44 MPa is required for the sequential stress-induced martensitic transformation. The difference of 0.5 K in $\Delta T_{ad,cyclic,EHC}$, though in 5 and 10 T the reverse martensitic transformation is fully induced (see Figure 12(a)) can result from a slight offset in $T_{start}$.

### 3.8. Comparison of caloric concepts

Figure 14 summarizes the cyclic effects of different caloric cooling concepts for the suction-cast Ni-Mn-In sample. The maximum $\Delta T_{ad,cyclic}$ can be achieved via elastocaloric cooling exhibiting -6.3 K when a stress of 300 MPa is applied and removed at 305 K. By the exploiting-hysteresis cycle a maximum cyclic effect of -4.9 K can be estimated when a magnetic field of 10 T and a sequential stress of 44 MPa are applied at 285 K. Compared to elastocaloric cooling, significantly lower stresses are required in the exploiting-hysteresis concept for the stress-induced martensitic transformation. The reasons for this are twofold: Firstly, the absolute temperature of the material is reduced in the exploiting-hysteresis concept. Secondly, the exploiting-hysteresis concept allows lower strain rates upon the stress induced martensitic transformation as the material is only set back to its initial martensite state without the necessity of achieving a $\Delta T_{ad}$ during the loading step. As a consequence of the reduced critical stresses the materials' lifetime will drastically improve in cyclic operation. It should be emphasized, that in the exploiting-hysteresis concept a large cyclic $\Delta T_{ad}$ of -4.1 K still can be estimated for moderate field changes of 1.9 T with a subsequent stress of 55 MPa. This corresponds to an increase of almost 37 % compared to the highest reported cyclic $\Delta T_{ad}$ of -3 K which could be achieved in Ni-Mn-based Heusler compounds for similar field changes by conventional magnetic cooling [72]. In the present suction-cast Ni-Mn-In sample a maximum cyclic effect of -1.2 K was obtained for a field change of 1.9 T.

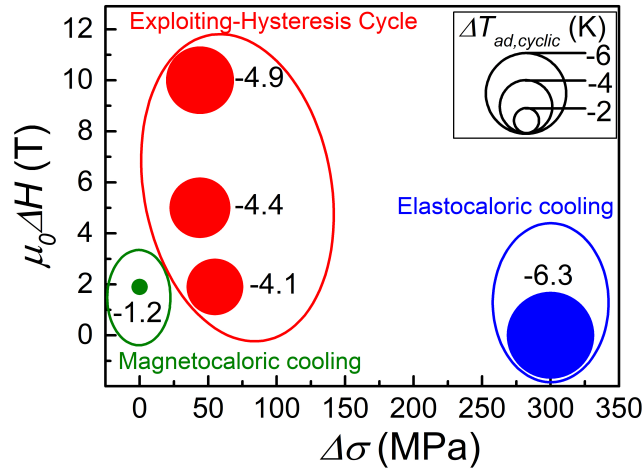

Figure 14: Cyclic adiabatic temperature change $\Delta T_{ad,cyclic}$ of suction-cast Ni-Mn-In for different caloric cooling concepts

## 4 Conclusions

In this study, we present a distinct influence of microstructure on the application of Ni-Mn-In Heusler compounds for multicaloric cooling using magnetic fields and uniaxial stress. The main findings are listed in detail below:

(1) Temperature-dependent x-ray powder diffraction reveals a martensitic transformation from high-temperature $L2_1$ austenite to low-temperature 3M monoclinic martensite. The lattice parameters and symmetry relation of the two phases were used to compute the strains of the stress-induced martensitic transformation under compression in various crystallographic directions. A topmost transformation strain of 5.8 % was found when the austenite phase is compressed in [001] direction.
(2) The comparison of suction-cast and arc-molten microstructures via SEM-EBSD revealed a <001> solidification texture for both synthesis techniques. By correlating these data with stress-strain measurements a strong effect of texture on the stress-induced martensitic transformation was demonstrated. It is shown that a <001> texture in compression direction is desriable to reduce the critical transformation stresses.
(3) The improvement of the compressive strength by grain refinement. By suction casting (grain diameter of 41 µm) an almost 40 % higher compressive strength was achieved than for conventional arc melting (grain diameter of 675 µm).
(4) In contrast to arc-molten Ni-Mn-In [45], an instantaneous response of the magnetic-field-induced reverse martensitic transformation was found for sweeping rates up to 1850 Ts$^{-1}$ in suction-cast Ni-Mn-In, which could result from a facilitated annihilation of retained martensite by a higher density of grain boundaries.

(5) Temperature-stress and temperature-magnetic field phase diagrams were established for suction-cast Ni-Mn-In. At higher external fields, an increasing thermal hysteresis and transition width was observed for both stimuli, even though uniaxial stress stabilizes the martensite and magnetic field the austenite phase.

(6) Isothermal entropy and adiabatic temperature changes were determined for the magnetic-field- and stress-induced phase transformation in suction-cast Ni-Mn-In. We illustrate that the maximum cyclic effect of -1.2 K in a magnetic field change of 1.9 T can be increased by more than 200 % to -4.1 K when a sequential stress of 55 MPa is applied. This significant adiabatic temperature change exceeds the topmost cyclic magnetocaloric effect reported for Ni-Mn-based Heusler in similar magnetic fields by more than one-third. In addition, the stresses required for the multicaloric approach are shown to be strongly reduced compared to elastocaloric cooling.

With this, we demonstrate that an optimized microstructural design and the combination of magnetic field and stress can enable large cyclic caloric effects in moderate external fields in the metamagnetic shape-memory alloy Ni-Mn-In. This finding should encourage the investigation of multicaloric effects in other first-order materials and clearly illustrates the potential of multi-stimuli concepts for environmentally-friendly refrigeration.

**Acknowledgements**

We thank Tom Keil for the support with the EBSD measurements. This work was funded by the European Research Council (ERC) under the European Union's Horizon 2020 research and innovation programme (grant no. 743116—project Cool Innov) and the CICyT (Spain), project MAT2016-75823-R. We acknowledge the HLD at HZDR, member of the European Magnetic Field Laboratory (EMFL) and the Helmholtz Association for the financial support via the Helmholtz-RSF Joint Research Group (Project No. HRSF-0045) and the Deutsche Forschungsgemeinschaft (DFG, German Research Foundation) via the Project-ID 405553726 – TRR 270 and the Project-ID GO 3343/2-1 – BEsT. A. G. acknowledges financial support from Universitat de Barcelona under the APIF scholarship.